\documentclass[aps,pra,preprint,a4paper,amsmath,amssymb,showpacs,%
floatfix,]{revtex4}

\usepackage{graphicx}
\usepackage{epstopdf}

\usepackage[usenames,dvipsnames]{color}

\begin{document}
%%%%%%%%%%%%%%%%%%%%%%%%%%%%%%%%%%%%%%%%%%%%%%%%%%
%
% Version of April 11, 2017
% 
%
%%%%%%%%%%%%%%%%%%%%%%%%%%%%%%%%%%%%%%%%%%%%%%%%%
\title{Vibrational-ground-state zero-width resonances for laser filtration: An extended semiclassical analysis.}

\date{\today} 

\author{Amine Jaouadi}
\affiliation{Qatar Foundation, P.O. Box 5825, Doha, Qatar }
\author{Roland Lefebvre}
\affiliation{ISMO, Univ. Paris-Sud, CNRS, Universit\'e Paris-Saclay, 91405 Orsay cedex, France}
\author{Osman Atabek}\email{osman.atabek@u-psud.fr}
\affiliation{ISMO, Univ. Paris-Sud, CNRS, Universit\'e Paris-Saclay, 91405 Orsay cedex, France}

%%%%%%%%%%      ABSTRACT   %%%%%%%%%%%%%%%%%%%%%%%%
\begin{abstract}
A semiclassical model supporting the destructive interference interpretation of zero-width resonances (ZWR) is extended to wavelengths inducing $c^-$-type curve crossing situations in Na$_2$	strong field dissociation. This opens the possibility to get critical couples of wavelengths $\lambda$ and field intensities $I$ to reach ZWRs associated with the ground vibrationless level $v=0$, that, contrary to other vibrational states ($v>0$), is not attainable for the commonly referred $c^+$-type crossings. The morphology of such ZWRs in the laser ($I, \lambda$) parameter plane and their usefulness in filtration strategies aiming at molecular cooling down to the ground $v=0$ state are examined within the frame of an adiabatic transport scheme.
\end{abstract}

\maketitle

%33.80.Gj Diffuse spectra; predissociation, photodissociation
%42.50.Hz Strong-field excitation of optical transitions in quantum systems; multiphoton processes; dynamic Stark shift
%37.10.Mn Slowing and cooling of molecules
%37.10.Pq Trapping of molecules

%31.15.p 	Calculations and mathematical techniques in atomic and molecular physics
%33.80.-b 	Photon interactions with molecules

%--------------------------------------------------------------------------------%

\section{Introduction}
\label{sec:intro}
Laser filtration based on zero-width resonances (ZWR) has already been referred to as a selective and robust technique that, starting from a given vibrational distribution, aims at shaping a chirped laser pulse such as to efficiently photodissociate all vibrational states, at the exception of but one \cite{catherine, AtabekRC}.
The basic mechanism is an adiabatic transport of the vibrational state to be filtrated (i.e., protected against dissociation) on its associated infinitely long-lived ZWR. Such control strategies have already been worked out for vibrational cooling purpose on the specific example of Na$_2$, prepared by photoassociation in some excited vibrational levels.
The theoretical observation is that for certain critical field parameters (wavelength $\lambda$ and intensity $I$) the photodissociation rate vanishes, resulting into a ZWR. The objective is to produce ZWRs at will and in a controllable way, continuously tuning laser parameters. A laser pulse is shaped in such a way to adiabatically transport a given vibrational state $v$ on its parent ZWR and to track it all along the pulse referring to an effective phase strategy \cite{Lecl2016}. For long enough durations, when the pulse is switched off, only the single vibrational state $v$ remains populated, all others $v'\neq v$ having decayed, leading thus to a robust vibrational cooling control strategy.

It has been shown that ZWRs result from destructive interference between two outgoing wave components  accommodated by laser-induced adiabatic potentials of a semiclassical two-channel description \cite{Bandrauk, child}. Roughly speaking, the critical phase matching of the interference scheme relies on the degeneracy of two energy levels: One which originates, in field-free conditions, from $v$  and the other $v_+$, supported by the field-dressed upper bound adiabatic potential. Depending on the wavelength, field dressing is such that a given $v$ could be brought in energy coincidence with any $v_+=0,1,...$, resulting into several ZWRs. But it has recently been argued that the ground vibrational level $v=0$ constitutes an exception (in a generic sense and at least for Na$_2$), without the possibility to merge into any ZWR \cite{Lef2011}. This does not however prevent a vibrational cooling objective, still achievable through filtration aiming at a single level protection $v'\neq 0$, although this is not the ground one. In a second step, a STIRAP process could then bring the $v'$ population onto $v=0$ \cite{guerin1998}.

The purpose of the present work is to extend the semiclassical analysis having in mind the specific goal of preparing the vibrationless state directly from a Boltzmann-type thermal distribution. This is done by referring to some wavelength regime inducing curve crossing schemes at internuclear distances less than the equilibrium geometry of $v=0$. Such situations are labeled as $c^-$ crossings, as opposite to  the $c^+$ ones  of previous investigations \cite{mulliken}.

The paper is organized in the following way: 
In section \ref{sec2},
ZWRs are introduced 
within a two-state photodissociation model of Na$_2$ in a time-independent close-coupled Floquet Hamiltonian formalism. Their interpretation is based on a semiclassical model with an original extension to $c^-$ avoided crossings. The filtration strategy based both on the ground and first excited vibrational states ($v=0,1$) are presented in Section \ref{sec4}.

\section{Photodissociation dynamics}\label{sec2}

%%%%%%%%%%%%%    PHOTODISSOCIATION MODEL      %%%%%%%%%%%%%%%%
\subsection{Zero-width resonances: Quantum description} \label{photodissmodel}

Rotationless Na$_2$ multiphoton dissociation is described within a two electronic states Born-Oppenheimer approximation. These are labeled $\arrowvert 1 \rangle$ for the bound a$^3\Sigma_u^+ (3^2S+3^2S)$ and $\arrowvert 2 \rangle$ for the dissociative excited state (1)$^3\Pi_g (3^2S+3^2P)$. $R$ being the internuclear distance, the nuclear components $\phi_{1,2}(R,t)$ of the time-dependent wave function:
\begin{equation}
 \arrowvert \Phi (R,t) \rangle = \vert \phi_1(R,t) \rangle \arrowvert 1 \rangle +
\vert  \phi_2 (R,t) \rangle \arrowvert 2 \rangle.
\end{equation}
are solutions of the Time Dependent Schr\"odinger Equation (TDSE) :
\begin{eqnarray}
& i\hbar \frac{\partial}{\partial t} \left[\begin{array}{c} \phi_1 (R,t)  \\
\phi_2 (R,t) \end{array}\right] 
 =   \left( T_N + \left[\begin{array}{c c} V_1(R)&0 \\
0&V_2(R) \end{array}\right] \right. & \nonumber \\ 
& - \left. \mu_{12}(R)\mathcal{E}(t)\left[\begin{array}{c c} 0&1 \\
1&0 \end{array}\right] \right)  \left[\begin{array}{c} \phi_1 (R,t) \\
\phi_2 (R,t) \end{array}\right] & 
\label{TDSE}
\end{eqnarray}
 where $T_N$ represents the nuclear kinetic energy. $V_1(R)$ and $V_2(R)$ are the Born-Oppenheimer potentials and $\mu_{12}(R)$ is the transition dipole between $\arrowvert 1 \rangle$ and $\arrowvert 2 \rangle$.
 The linearly polarized electric field $\mathcal{E}(t)$ is given, for a continuous wave (cw) laser by:  
 \begin{equation}
 \mathcal{E}(t) = E\cos(\omega t)
 \end{equation}
The intensity and the wavelength are given by $I\propto E^2$ and $\lambda= 2 \pi c /\omega$, $c$ being the speed of light.
Due to time-periodicity, the Floquet ansatz leads to \cite{atabek03}:
\begin{eqnarray}
\left[ \begin{array}{c} \phi_1 (R,t) \\
\phi_2 (R,t) \end{array}\right] = e^{-iE_v t/\hbar} \left[ \begin{array}{c} \chi_1 (R,t)  \\
\chi_2 (R,t) \end{array}\right].
\label{eq:anzats}
\end{eqnarray}
where Fourier expanded $\chi_k(R,t)   (k=1,2)$:
\begin{equation}
\chi_k(R,t)=\sum_{n=-\infty}^{\infty} e^{in\omega t} \varphi_{k,n}(R)
\label{eq:expansion}
\end{equation}
involve components satisfying a set of coupled differential equations, for any $n$, which for moderate field intensities (retaining only $n=0,1$) reduce to: 
\begin{equation}
\left[ T_N +
V_1(R)+\hbar \omega-E_v \right]\varphi_{1,1}(R)-1/2 E \mu_{12}(R)
\varphi_{2,0}(R)=0 
\nonumber
\end{equation}
\begin{equation}
\left[ T_N +
V_2(R)-E_v \right]\varphi_{2,0}(R)-1/2 E \mu_{12}(R)
\varphi_{1,1}(R)=0   
\label{eq:closecoupled}
\end{equation}
Resonances are quantized solutions with Siegert type outgoing-wave boundary conditions \cite{siegert} and have complex quasi-energies of the form 
$\Re(E_v)-i\Gamma_v/2$, where $\Gamma_v$ is the resonance width related to its decay rate.
In the following, label $v$ designates both the field-free vibrational level and the laser-induced resonance originating from this vibrational state.

Going beyond the cw laser assumption, we consider 
a chirped laser pulse with parameters $\epsilon(t)\equiv\{E(t), \omega(t)\}$ involving slowly varying envelope and frequency. 
The purpose of optimizing laser parameters such that the survival probability of a resonance state originating in field-free conditions from a given vibrational state $v$ be maximized, while all other resonances (originating from $v'\neq v$) are decaying fast, is conducted within the frame of the adiabatic Floquet formalism \cite{guerin}.
The full control strategy consists in trapping the system into a single eigenvector of the adiabatic Floquet Hamiltonian, in a so called extended Hilbert space and shaping a pulse with field parameters such that this eigenstate presents the lowest (zero, if possible) dissociation rate. We have recently shown \cite{Lecl2016} that this is achieved by an optimal choice for the field parameters:
\begin{equation}
\epsilon^*(t)\equiv\{E^{ZWR}(t), \omega_{\text{eff}}^{{ZWR}}(t)\}
\nonumber 
\end{equation}
$\omega_{\text{eff}}$ being an effective frequency in this extended Hilbert space and
such that:
\begin{equation}
\Im[E_v\{\epsilon^{ZWR}(t)\}]=0 ~~ \forall t
\label{eq:imag}
\end{equation}
where $\Im(E_v)$ is the imaginary part of the energies of these field-induced resonances.  
Eq.(\ref{eq:imag}) is nothing but a ZWR path (originating from $|v \rangle$) in the amplitude, frequency parameter plane (or equivalently, intensity $I$, wavelength $\lambda$) as a function of $t$, that is:
\begin{equation}
\epsilon^{\text{ZWR}}(t)\equiv
\left\{
\begin{array}{c}
E^{\text{ZWR}}(t) = [I^{\text{ZWR}}(t)]^{1/2}, \\ 
\omega_{\text{eff}}^{\text{ZWR}}(t) = 2\pi c/\lambda^{\text{ZWR}}(t)  
\end{array}
\right\}
\end{equation}
Finally, the optimal laser pulse acting in the original Hilbert space where the evolution is monitored by the TDSE displayed in Eq.(\ref{TDSE}) is given by \cite{Lecl2016}:
\begin{equation}
\mathcal{E}^*(t)= [I^{\text{ZWR}}(t)]^{1/2} \cdot  \cos \left({\int_0^t 2\pi c/\lambda^{\text{ZWR}}(t') dt'} \right)
\label{adiab_pulse}
\end{equation}
It has in particular been shown that ZWRs are good candidates for a full adiabatic Floquet treatment as initially derived for pure bound states \cite{guerin, Lecl2016}.
The molecule, initially in a particular field-free vibrational state $v$, is supposed to be adiabatically driven by such a pulse. Adiabaticity means here that a single resonance $\Phi_v(t)$, labeled $v$ according to its field-free parent bound state, is followed during the whole dynamics. This resonance wave function involves, through its complete basis set expansion, a combination of both bound and continuum eigenstates of the field-free molecular Hamiltonian. But the important issue is that, at the end of the pulse, the molecule is again on its initial single vibrational state $v$ (adiabaticity condition). For such open systems, contrary to dynamics involving  bound states only, there is unavoidably an irreversible  decay process, precisely due to the fact that vibrational continuum states are temporarily populated under the effect of the pulse, even though this is minimized by a ZWR path tracking. A quantitative measure of such a decay is given in terms of the overall fraction of non-dissociated molecules, assuming a perfect adiabatic following of the selected resonance
 \cite{catherine}:
\begin{equation} \label{P-undiss}
 P_{v}(t) \; = \; \exp\left[- \hbar^{-1}\int_0^{t} \Gamma_v(\epsilon (t'))~dt'\right] \;\; 
\end{equation}
where the decay rate $\Gamma_v(\epsilon(t))$ is associated with the relevant Floquet resonance quasi-energy $E_v(\epsilon(t))$ using the instantaneous field parameters $ \epsilon(t) \equiv\{E(t), \omega(t)\}$ at time $t$.

The control issue consists in investigating how rates are changing with the field parameters and in particular find optimal combinations $\epsilon^{\text{ZWR}}(t)$ for which these rates are small enough (or even ideally zero) to insure the survival of the vibrational state $v$ to the laser excitation, that is $P_{v}(\tau) \approx 1$, $\tau$ being the total pulse duration.

%%%%%%%%%%%%%%%%%   SEMICLASSICAL MODEL   %%%%%%%%%%%%%%%%%%

\subsection{Zero-width resonances: Semiclassical model}

\begin{figure}
	\includegraphics[width=1.4\linewidth]{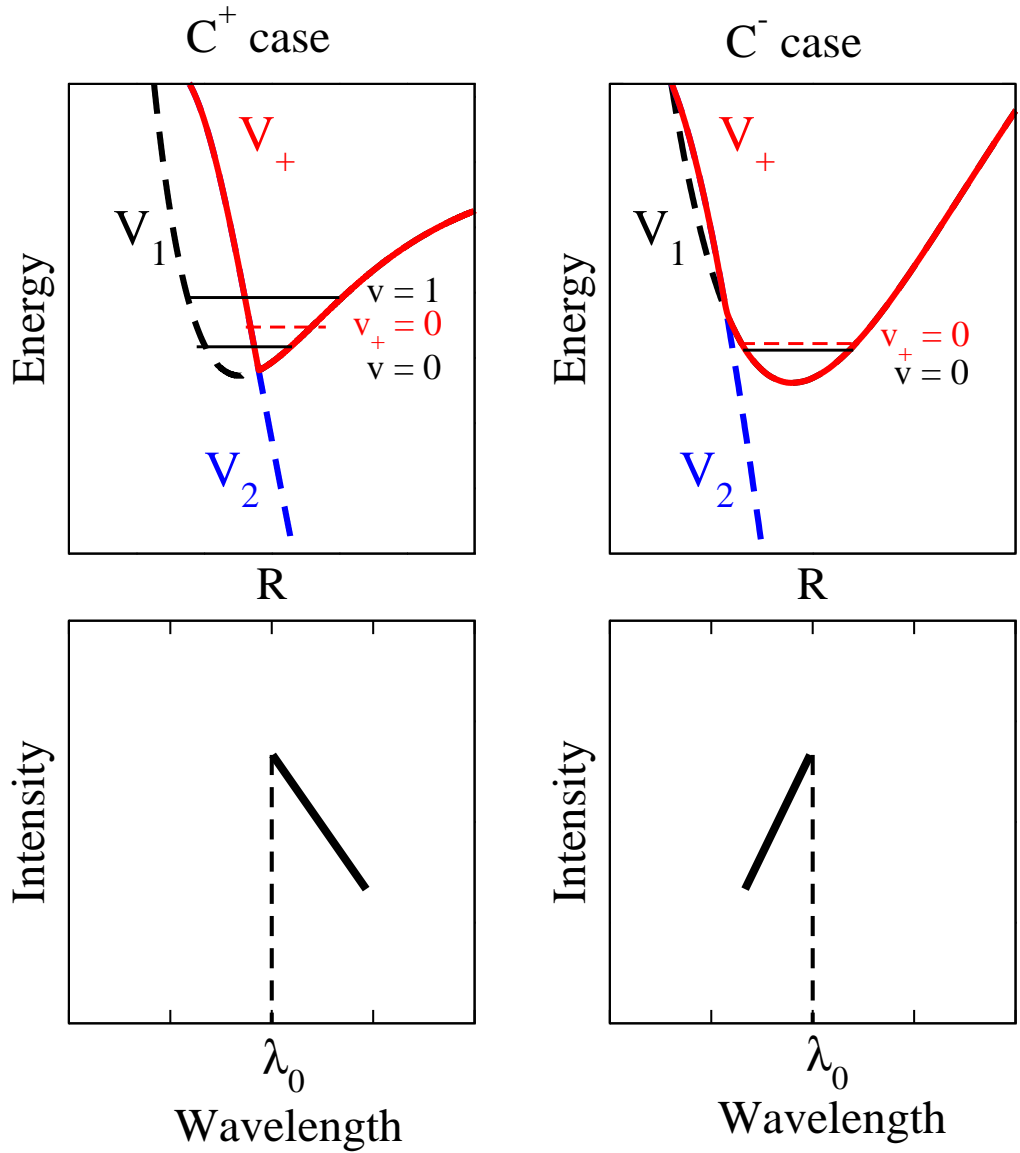}
	\caption{Schematic view of field-dressed potential energy curves (upper panels) and ZWR maps (lower panels) for the $c^+$ (left panels) and the $c^-$ (right panels) cases. Na$_2$ upper adiabatic potentials $V_+(R)$ for negligible field intensities are indicated by solid red lines.}
	\label{schematic_pot}
\end{figure}

A full destructive interference interpretation of ZWRs is provided by a semiclassical theory based on a two-channel scattering model, involving adiabatic potentials $V_{\pm}(R)$ resulting from the diagonalization of  molecule-field interaction matrix \cite{Bandrauk,Child2}. Dissociation quenching related to a null value of the outgoing scattering amplitude in the lower (open) adiabatic channel $V_-$ occurs when the following two conditions are simultaneously fulfilled \cite{Bandrauk,Child2}:
\begin{equation}
	\int_{R_{+}}^{R_c} dR'~k_{+}(R')+\int_{R_{c}}^{R_{t}} dR' k_{+}(R')+ \chi=(\tilde{v}_{+}+\frac{1}{2})\pi
	\label{eq:semiclassical1}
\end{equation}
and
\begin{equation}
	\int_{R_{-}}^{R_c} dR'~k_{-}(R')+\int_{R_{c}}^{R_{t}}dR' k_{+}(R') =(\tilde{v}+\frac{1}{2})\pi
	\label{eq:semiclassical2}
\end{equation}
The wavenumbers are given by: 
\begin{equation}
k_{\pm}(R)=\hbar^{-1}[2m(\varepsilon-V_{\pm}(R)]^{1/2}.
\label{eq:wavenumbers}
\end{equation}
$m$ is the reduced nuclear mass, $R_{\pm}$ are the left turning points of $V_{\pm}$ potentials, $R_t$ is the right turning point of $V_+$ and $R_c$ is the diabatic crossing point resulting from field-dressing. With integer quantum numbers $\tilde{v}_+$ and $\tilde{v}$ these conditions are nothing but the requirements of Bohr-Sommerfeld quantization involving a coincidence between two energies; namely one $\varepsilon=\varepsilon_{\tilde{v}^+}$ of the upper adiabatic potential $V_+(R)$, with a phase correction $\chi$, which in weak coupling is $-\pi/4$ \cite{Child2}, and another $\varepsilon=\varepsilon_{\tilde{v}}$ of a potential $\tilde{V}(R)$ made of two branches, namely, $V_-(R)$ for $R \leq R_c$, and $V_+(R)$ otherwise. For a weak coupling, this is practically the field-free diabatic attractive potential $V_1(R)$. An analytical expression of the resonance width $\Gamma_v$ is \cite{Child2}:
\begin{equation}
	\Gamma_v= \frac{2\pi}{\hbar}  \frac{e^{2\pi\nu} (e^{2\pi\nu}-1) \omega_d \omega_+}{[\omega_++(e^{2\pi\nu}-1)\omega_d]^3} (\varepsilon_{\tilde{v}}-\varepsilon_{\tilde{v}_{+}})^2
	\label{eq:Gammasem}
\end{equation}
This expression clearly displays the role played by such energy coincidences, in terms of the square of their differences.
In Eq.(\ref{eq:Gammasem}) 
$\omega_d$ and $\omega_+$ are the local energy spacings of the modified diabatic and adiabatic potentials respectively. $\nu$ is the Landau-Zener coupling parameter:
\begin{equation}
	\nu=\frac{\mu_{12}^2(R_c)E^2}{\hbar\bar{v}|\Delta F|}
	\label{eq:Landau-Zener}
\end{equation}
where $\bar{v}$ and $\Delta F$ are the classical velocity and slope difference of the diabatic potentials at $R_c$.
Clearly, the two energies $\varepsilon=\varepsilon_{\tilde{v}}$ and $\varepsilon=\varepsilon_{\tilde{v}^+}$, and therefore the width $\Gamma_v$ are dependent on field parameters, i.e., both frequency (or wavelength) and amplitude (or intensity). This is in particular due to the ($\lambda$, $I$)-dependence of the corresponding field-dressed adiabatic potentials $V_{\pm}(R)$. As a consequence, ZWRs in photodissociation can be produced at will by a fine tuning of the wavelength and intensity.
 Moreover, for a wavelength $\lambda$ which roughly brings into coincidence the levels $\tilde{v}$ (corresponding to the field-free vibrational level $v$ in consideration) and $\tilde{v}_+=0$, a fine tuning of the intensity $I$ will result in an accurate determination of a ZWR, that is $\Gamma_v(\lambda , I)=0$. 
In some cases, a stronger field (higher $I$) may also bring into coincidence $\tilde{v}$ with $\tilde{v}_+=1$, producing thus a second ZWR, for the same wavelength, and so on for $\tilde{v}_+=2,3...$ But, one can also envisage slightly different wavelengths which build energetically close enough $\tilde{v}$ and $\tilde{v}_+$ levels in a field-dressed picture, such that a subsequent fine tuning of the intensity brings them into precise coincidence. This flexibility offered by the field parameters that, in principle, can be continuously modified, is at the origin of not only quasi-zero width photodissociation resonances, but also for their multiple occurrence in the ($\lambda$, $I$)-parameter plane \cite{multiZWR}. 

We emphasize that the semiclassical description of photodissociation is based on field-dressed adiabatic potential energy curves with avoided crossings mainly controlled by frequency, whereas couplings are intensity dependent. In addition, according to Child's diagrammatic approach we are following here, these crossings should be reached classically for both channels, their turning points being at the left of the crossing point. Such a situation is the one which is valid for the most commonly refereed $c^+$-type crossings. But, as will be discussed hereafter, the semiclassical analysis could still be useful, for some  $c^-$-type crossings, occurring at higher frequencies, compatible with the classically allowed picture and leading to the previously discarded possibility to reach a ZWR associated with $v=0$. Finally, it is worthwhile noting that some extensions, at even higher frequencies, to classically non-reachable crossing situations have already been worked out using complex crossing points \cite{atabeklefebvre}. 

We now examine in more detail some generic properties of ZWR behaviors in the ($\lambda, I$) parameter plane, by distinguishing the $c^+$ and $c^-$ cases.

%%%%%%%%%%%%%%%%%%%%     THE c+ CASE    %%%%%%%%%%%%%%%%%
\subsubsection{Low frequency regime: $c^+$-type crossing.}

Such cases correspond to low frequency field-dressing with a diabatic curve crossing at the right of the equilibrium distance, $R_c > R_e$. A typical situation is the one illustrated on the left panel of the schematic view of Fig.\ref{schematic_pot} with the lowest possible wavelength $\lambda_0$ leading to $R_c = R_e$. 
Semiclassical rationalization of ZWRs generic behaviors, according to the energy coincidences between $\varepsilon_{\tilde{v}}$ and $\varepsilon_{\tilde{v}_{+}}$ involved in Eq.\ref{eq:Gammasem}, can be conducted in three steps: (i) field-dressing with $\lambda$ (intensity being taken as negligible), which is the major shifting effect on $v_+$ levels not affecting $\tilde{v}=v$; (ii) introduction of the additional phase $\chi$ (taken as $-\pi/4$, for low enough intensities) affecting $v_+$ which becomes $\tilde{v}_{+}$; (iii) consideration of the role played by the field intensity in locally changing the adiabatic potentials supporting both $\tilde{v}$ and $\tilde{v}_{+}$.

The argument discarding the possibility of a ZWR associated with $v=0$ is based on the fact that in a field-dressed picture the upper adiabatic potential $V_+(R)$ accommodating level $v_+$, presents a local  curvature (close to $R_c$) higher than the one of the bound diabatic state $V_1(R)$ supporting level $v$, at least for low intensities.
This is depicted in Fig.\ref{schematic_pot} for the lowest possible wavelength $\lambda = \lambda_0$, most favorable candidate for an energy coincidence. The consequence is that
\begin{equation}
\varepsilon_{\tilde{v}_{+}=0} > \varepsilon_{\tilde{v}=0}.
\label{eq:coincidence}
\end{equation}
Moreover, the phase $\chi$ produces an additional energy increase on $\varepsilon_{\tilde{v}_{+}=0}$. Finally, the effect of the field intensity is such that it will affect the system by increasing $\varepsilon_{\tilde{v}_{+}=0}$, while slightly decreasing $\varepsilon_{\tilde{v}=0}$.
Obviously, all other wavelengths ($\lambda > \lambda_0$) of this low frequency $c^+$ regime will shift $\varepsilon_{\tilde{v}_{+}=0}$  at even higher energies. The coincidence condition for $v=0$, required by Eq.\ref{eq:Gammasem}, can never be fulfilled for wavelengths inducing a $c^+$-type crossing.

For all other levels $v > 0$, semiclassical expectations are different. As is clear from Fig.\ref{schematic_pot}, for $\lambda = \lambda_0$ and $v=1$,
\begin{equation}
\varepsilon_{\tilde{v}_{+}=0} < \varepsilon_{\tilde{v}=1}.
\label{eq:coincidence2}
\end{equation}
Both the neglected additional phase $\chi$ and changes in field control parameters (increase of wavelength $\lambda > \lambda_0$ and intensity $I$) result in increasing $\varepsilon_{\tilde{v}_{+}}=0$. It is important to note that $\varepsilon_{\tilde{v}=1}$ is in turn affected by the increase of the field intensity, but much less than $\varepsilon_{\tilde{v}_{+}=0}$. Specific laser parameters ($\lambda^{ZWR}, I^{ZWR}$) could then be found to achieve the semiclassical energy coincidence of Eq.\ref{eq:Gammasem}, leading to ZWR($v=1, v_+=0$) originating from $v=1$.
	
To follow a typical ZWR map in the laser parameter plane, we suppose that a first coincidence ($\varepsilon_{\tilde{v}} = \varepsilon_{\tilde{v}_{+}}$) has been obtained for some critical ($I, \lambda$) parameters. When the wavelength is progressively increased, $\varepsilon_{\tilde{v}_{+}}$ is blue shifted, whereas $\varepsilon_{\tilde{v}}$ is only slightly affected. The coincidence required for a ZWR is no more achieved. In order to compensate the increase of $\varepsilon_{\tilde{v}_{+}}$ we have to lower the field intensity. As a consequence, the ZWR path in the ($I, \lambda$)-plane is of negative slope, as illustrated in Fig.\ref{schematic_pot}.

%%%%%%%%%%%%%%%%%%%%     THE c+ CASE    %%%%%%%%%%%%%%%%%
\subsubsection{High frequency regime: $c^-$-type crossing.}

%%%%%%%%%%%%%%%%%%%%%%%%%%%%%%%%%%%%%%%%%%%%%%%%%%%%%%

At higher frequencies, for $\lambda < \lambda_0$, curve crossings occur on the left of the equilibrium geometry, that is $R_c < R_e$. The right panels of Fig.\ref{schematic_pot} illustrate such a typical situation. As already mentioned, there is still a wavelength window for which, even in this regime, the semiclassical model is still valid, with left turning points of Eq. \ref{eq:Gammasem} satisfying the following condition: 
\begin{equation}
R_- < R_+ < R_c < R_0.
\label{eq:Rinequal}
\end{equation}
\begin{figure}
	\includegraphics[width=0.8\linewidth]{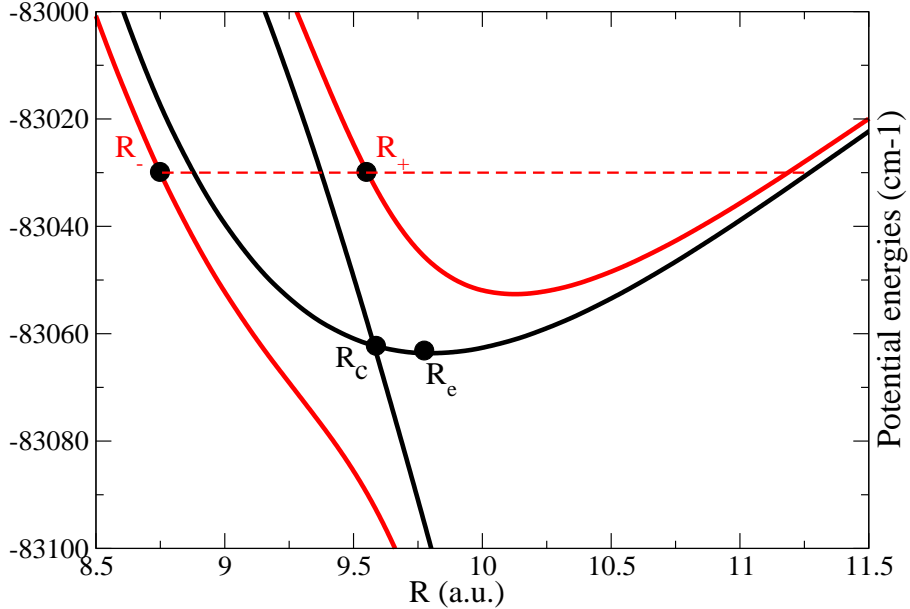}
	\caption{Potential energy curves of the a$^3\Sigma_u^+ (3^2S+3^2S)$ triplet (bound) and (1)$^3\Pi_g (3^2S+3^2P)$ excited (dissociative) electronic states of Na$_2$ (solid black). Adiabatic potential energies for a $c^-$ case corresponding to a wavelength $\lambda$=551.5 $nm$ and an intensity $I=0.468 \times 10^9 W/cm^2$ are given in solid red, together with relevant internuclear distances ($R_e$ and $R_c$ for the equilibrium and crossing positions, $R_-$ and $R_+$ for the left turning points).}
	\label{adiabatic_pot}
\end{figure}
At negligible field intensities, the two laser-dressed potentials $V_1(R)$ and $V_+(R)$ are very similar, at least in the deeper part of their common well accommodating the lowest vibrational levels, and in particular $v=0$. The consequence is that the coincidence condition could be reached, at least approximately. It is important to note that, even if for $v > 0$ such $c^-$-type crossings are merely an extension of the wavelength regime for which ZWRs are expected, the situation is completely different for $v=0$. Concerning semiclassical energy coincidence arguments, this actually appears to be the only regime where, at least approximately, one can expect $\varepsilon_{\tilde{v}_{+}=0} \simeq \varepsilon_{\tilde{v}=0}$, that is to get a ZWR originating from the vibrationless ground state $v=0$ at field free conditions.

Finally, similarities between adiabatic potentials $\tilde{V}(R)$ and $V_+(R)$ regarding their common potential well around equilibrium geometry, are better obtained by: (i) decreasing the wavelength $\lambda < \lambda_0$ in order to enhance the potential well extension; (ii) decreasing the intensity $I$ in order to reduce energy separations of the avoided crossing area. The conclusion is that ZWRs paths in the ($I, \lambda$)-plane should behave with a positive slope, as opposite to the $c^+$-type crossing situation.

%%%%%%%%%%%%%%%    RESULTS AND DISCUSSION   %%%%%%%%%%%%%
\section{Results} \label{sec4}

Transitionally and rotationally cold, tightly bound and vibrationally hot Na$_2$ species are experimentally produced by photoassociation in the metastable bound state $^3\Sigma_u^+ (3^2S+3^2S)$, considered as an initial ground state (referred to as state 1) radiatively coupled with the repulsive, thus dissociating excited (1)$^3\Pi_g (3^2S+3^2P)$ electronic state (referred to as state 2). The corresponding Born-Oppenheimer potential energy curves $V_{1,2}(R)$ and the electronic transition dipole moment $\mu_{12}(R)$ between states 1 and 2 are taken from the literature \cite{magnier,aymar, AtabekPRL}. Finally, Na$_2$ reduced mass is taken as 20963.2195 $au$. As depicted in Fig.\ref{adiabatic_pot}, the equilibrium geometry corresponds to $R_e$ = 9.79 $au$. The critical wavelength for $c^0$ crossing (for $R_c=R_e$) is actually $\lambda_0=552$ $nm$, such that $c^+$-type crossings are obtained for $\lambda > \lambda_0$, whereas wavelengths $\lambda < \lambda_0$ lead to $c^-$-type.
Moreover, the lowest possible wavelength still fulfilling the requirement of Eq. \ref{eq:Rinequal}, for $v=0$, turns out to be $\lambda=550$ $nm$. This means that the $c^-$ extension of the semiclassical model within its diagrammatic presentation would only concern a moderate range of wavelengths, namely 550 $nm < \lambda < 552$ $nm$. 

Photoassociation typically prepares vibrational levels with quantum numbers $v \geq 8$. We have previously used a filtration strategy using ZWR tracking in the ($I,\lambda$) parameter plane by adiabatically transporting $v=8$ level on its associated ZWR \cite{AtabekRC}. This leads, after the laser pulse is over, to vibrational population left only on $v=8$ and therefor achieves efficient and robust cooling by preparing a single vibrational level, although not the ground one $v=0$. In other experimental situations, with an initial thermal distribution of vibrational states, the filtration targeting the vibrationless ground state $v=0$ would require the generalization of a similar strategy but now based on a ZWR associated with $v=0$. In the following, we start with the more common case of filtration referring to ZWR($v=1$) to illustrate both $c^+$ and $c^-$-type crossing behaviors, in conformity with the previous semiclassical model. In a second attempt, we analyze the case of ZWR($v=0$), showing that, with specific range of wavelengths (roughly inducing $c^-$-type crossings), efficient filtration still remains possible. 
%%%%%%%%%%%%%%%%%%%%     ZWRs MORPHOLOGY    %%%%%%%%%%%%%
\subsection{Filtering using ZWR($v=1$).} \label{zwr1}

\begin{figure}
	\includegraphics[angle=0,width=1.0\linewidth]{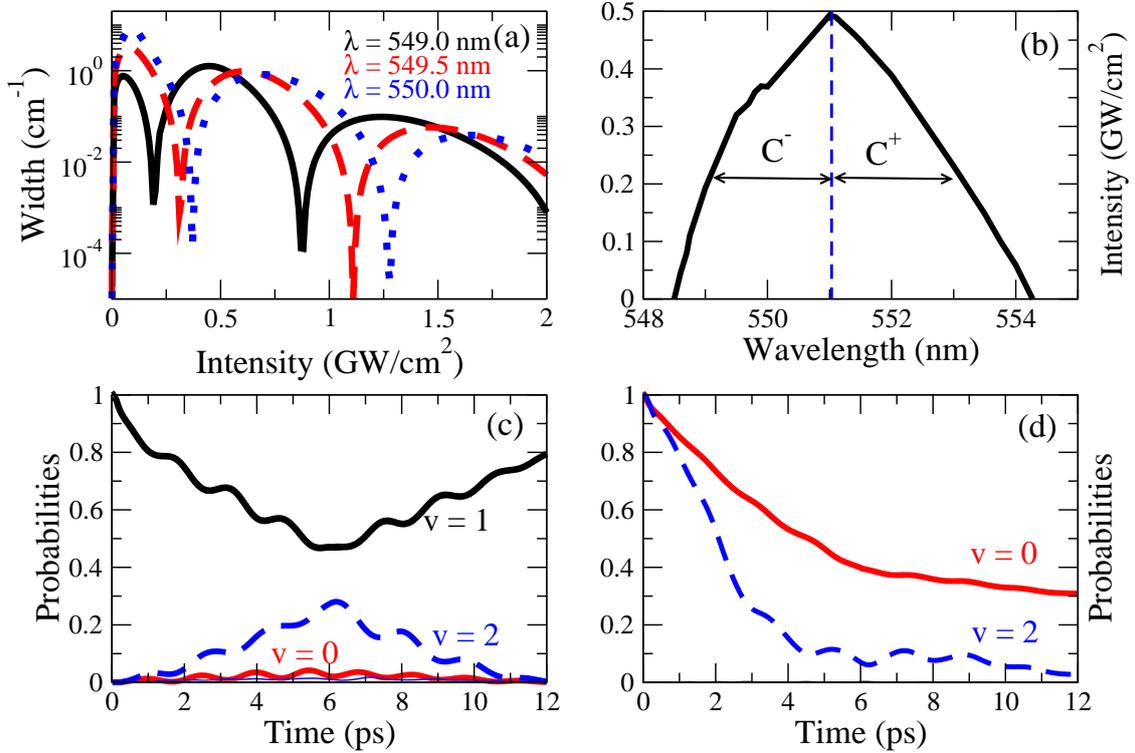}
	\caption{ZWR($v=1$) morphology and filtration strategy. Panel (a): 
Resonance widths as a function of intensity for selected wavelengths 
($\lambda$=549 $nm$, 549.5 $nm$, 550 $nm$) in log scale. Panel (b): ZWR map in the 
($I, \lambda$) parameter plane; Panel (c) Vibrational populations as a function 
of time with initial state $v=1$; Panel (d) Vibrational populations as a 
function of time with initial states $v=0,2$.}
	\label{fig:zwr1}
\end{figure}

Solving time-independent coupled equations Eq.(\ref{eq:closecoupled}) with 
Siegert boundary conditions for a set of continuous wave cw laser parameters 
$\{I,\lambda\}$, gives rise to resonances with complex eigenvalues $E_v$ 
correlating, in field-free conditions, with the real vibrational eigenenergies. 
We are actually interested in finding specific couples of field 
parameters for which the imaginary part of resonance eigenvalues are close to 
zero. More specifically, we analyze the wavelength regime 549 $nm < \lambda < 
\lambda_0$ corresponding to the semiclassical extended $c^-$-type crossing 
region for $v=1$.

Results of exploratory calculations illustrating the behavior of resonances originating from $v=1$ are displayed in panel (a) of Figure \ref{fig:zwr1}. We have selected three wavelengths within the semiclassical extension window and intensities up to $I = 2 GW/cm^2$. The overall tendency is a smooth regular decrease of the widths for increasing field strengths, in agreement with the generic behavior of Feshbach-type resonances (due to decreasing non-adiabatic couplings \cite{chrysos}). But, more interestingly, for specific intensities, we obtain sharp dips corresponding to resonance widths typically less than $10^{-3} cm^{-1}$, clear signatures of ZWRs. Within numerical inaccuracies inherent to the evaluation of such very small width resonances  (less than $10^{-6} cm^{-1}$), we observe that there are several couples of critical wavelengths and intensities producing ZWRs originating from a single vibrational level $v$. Figure \ref{fig:zwr1}, panel b, displays in the ($I, \lambda$) laser parameter plane, ZWRs path originating from ($v=1$) for both $c^-$ and $c^+$ regimes. As expected from the extended semiclassical analysis summarized in Fig.\ref{schematic_pot}, the $c^-$-type crossing region reached for 549 $nm < \lambda < \lambda_0$, roughly corresponds to a ZWR path with a positive slope. This is to be contrasted with the behavior in the $c^+$ crossing region $\lambda > \lambda_0$, where the slope is negative, once again in conformity with the semiclassical analysis of Fig.\ref{schematic_pot}. 
The last step for the filtration control is to shape frequency chirped laser pulses resulting from the effective phase adiabatic transport strategy of Eq.(\ref{adiab_pulse}),
where $\lambda^{ZWR}$ and $I^{ZWR}$ are those depicted in panel (b), exclusively for the $c^-$ region.
A wavepacket evolution based on TDSE solved by a third-order split-operator technique \cite{AtabekRC, feit}, gives the vibrational population dynamics under the effect of such a pulse acting either on $v=1$ as an initial state, or neighboring $v=0,2$ levels. The results are displayed in Fig.\ref{fig:zwr1} in panels c and d. The efficiency of filtration strategy is well proven. The vibrational population of level $v=1$ is well protected against dissociation (up to 80$\%$, on panel c), whereas the neighboring levels populations are decaying fast (panel d). We emphasize that similar observations have previously been discussed for $c^+$-type crossings, the originality of the present work is to show their possible extension to $c^-$-type crossings. We however notice that when referring to ZWRs in the $c^-$ semiclassical extension regime, the filtration process is slightly less selective (remaining $v=0$ population being not less than 30$\%$). This is due to the fact that ZWRs originating from $v=1$ and $v=0$ are close to each other. As is clear from Figure \ref{schematic_pot}, laser parameters inducing the energy coincidence which is looked for $v=0$ approximately correspond to the ones valid for a similar coincidence for $v=1$.

%%%%%%%%%%%%%%%%%%%%%%%%%%%%%%%%%%%%%%%%%%%%%%%%%%%%%%%%%%%%%%%%%%%
\subsection{Filtering using ZWR($v=0$).} \label{zwr0}

Having shown the validity of a possible extension of the semiclassical approach to $c^-$-type crossings, we are now in a position to examine the most challenging case of a ZWR originating from the vibrationless ground ($v=0$) state together with its potentiality to support robust filtration control.
\begin{figure}
	\includegraphics[angle=0,width=1.0\linewidth]{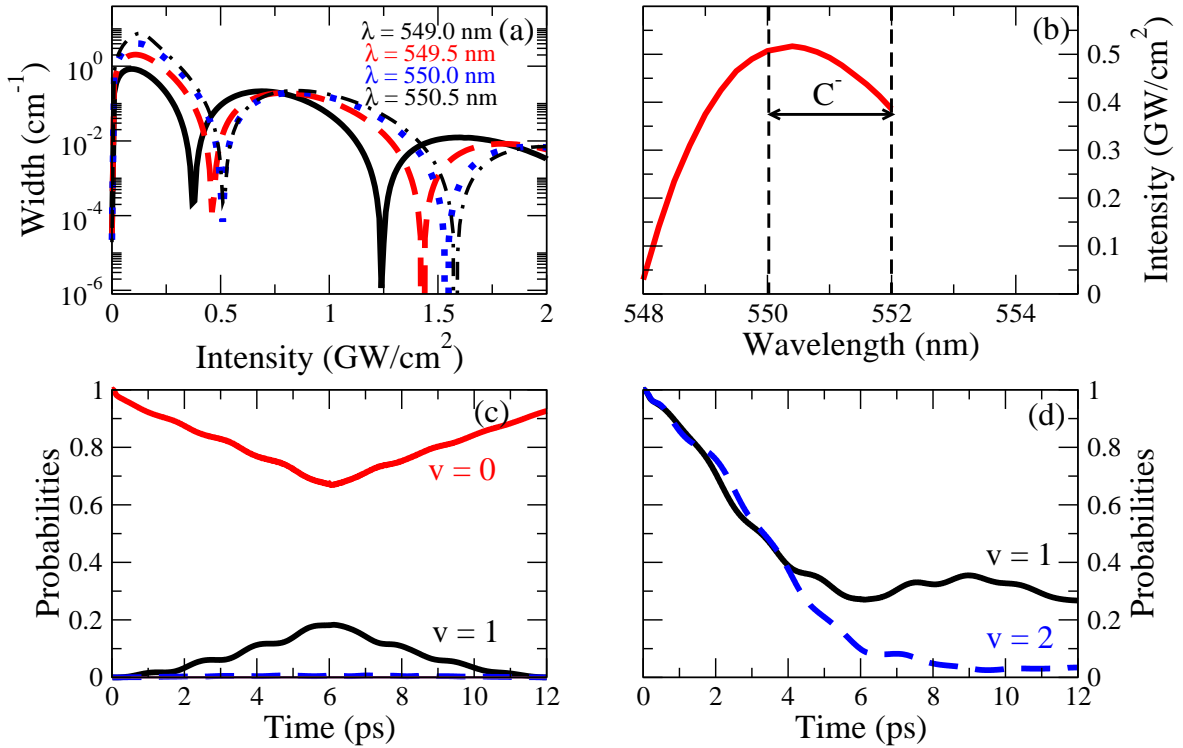}
	\caption{Same as for figure 3, but for ZWR($v=0$) and selected wavelengths $\lambda$= 549 $nm$, 550 $nm$, 551 $nm$ and 552 $nm$.}
	\label{fig:zwr02}
\end{figure}
The results are gathered in Fig.\ref{fig:zwr02} following a graphical illustration similar to the above discussed case of $v=1$. As previously, exploratory calculations are carried out for $c^-$-type crossing regime, covered by 550 $nm < \lambda <\lambda_0$. A selection of three such wavelengths is illustrated in panel a for the resonance widths originating from $v=0$ as a function of intensity. This clearly shows the possibility to reach ZWRs as sharp dips (widths typically less than $10^{-4} cm^{-1}$) superimposed to a smoothly decreasing background. More unexpectedly, some smaller wavelengths are also producing ZWRs. One of them corresponding to $\lambda = 549$ $nm$ is shown in Fig.\ref{fig:zwr02}. Panel b displays the ZWR($v=0$) path in the ($I, \lambda$) parameter plane. A few observations deserve interest: (i) No ZWR is obtained in the low frequency regime, for wavelengths $\lambda > \lambda_0$ leading to $c^+$-type crossing, in conformity with the semiclassical analysis; (ii) Unexpectedly, for wavelengths $\lambda < 550$ $nm$, ZWR are still observed, although the semiclassical model is no more valid, due to the fact that the crossing is not within the classically allowed region (or even no crossing at all); (iii) In the intermediate wavelength regime 550 $nm < \lambda < \lambda_0$ fully supported by the extended semiclassical $c^-$-type crossing, the slope of ZWRs path is positive in a region well on the left of $R_e$ (550 $nm < \lambda < 550.5$ $nm$) as expected from the analysis of Fig.\ref{schematic_pot}. But, when $R_c$ becomes closer to $R_e$, the slope changes to be negative, presumably due to a competition between decreasing energy separations ($\varepsilon_{\tilde{v}_{+}=0} - \varepsilon_{\tilde{v}=0}$) on the one hand, and increasing additional phase $\chi$ on the other hand, when the field strength is decreasing. Panel c shows the robustness of $v=0$ population efficiently protected against dissociation (up to 95$\%$), whereas panel d displays populations of neighboring states ($v=1,2$) which are decaying fast, but with still 27$\%$ remaining $v=1$ population at the end of the pulse, for reasons similar to those already discussed in the previous paragraph.

%%%%%%%%%%%%%%%%%%%%%%%%%%%%%%%%%%%%%%%%%%%%%%%%%%%%%%%%%%%%%%%%%%%%%%%%%%%%%%%%%%%%%%%%%%%
\section{Conclusion}\label{sec5}

The diagrammatic semiclassical model, of crucial importance in the destructive interference interpretation of ZWRs and in their localization in the laser ($I, \lambda$) parameter plane, is extended to wavelengths inducing 
$c^-$-type crossings in adiabatic potentials description. Such an extension remains however limited to wavelengths windows of moderate size, as additional requirements of classically reachable crossings within vibrational wavefunctions spatial stretching should be fulfilled. With this extension, the validity of which is first checked on the standard case of $v=1$, the semiclassical model acquires the capacity of a possible depiction of ZWRs originating, in field-free conditions, from the vibrationless ground state $v=0$, by approximately defining a couple of ($I, \lambda$) parameters. Actually, quantum Floquet photodissociation theory confirms these ZWR($v=0$) parameters by refining their values. But more unexpectedly, additional ZWRs($v=0$) are obtained in the $c^-$ regime, even though classical conditions are no more fulfilled.

A full quantum wave packet propagation shows that an adiabatic transport of population from $v=0$ to its associated ZWRs($v=0$) tracked all along an appropriately shaped laser pulse duration results in efficient vibrational population protection against photodissociation. When the pulse is over, the $v=0$ population is almost unchanged, pointing thus to the robustness of the mechanism. At the same time, all other ($v > 0$) vibrational populations of the initial thermal distribution are decaying, pointing to the selectivity of the filtration process, even though this is less than the one of the $c^+$ case.

As a conclusion, a laser controlled filtration strategy based on ($v=0$) ZWR tracking is shown to be robust and selective enough for molecular vibrational cooling aiming at obtaining the ground vibrational state in a single step laser excitation.

%%%%%%%%%%%%%%%%%%%%%%%%%%%%%%%%%%%%%%%%%%%%%%%%%%%%%%%%%%%%%%%%%%%%%%%%%%%%%%%%%%%%%%%%%%
%%%%%%%%%%%%%%%%    ACKNOWLEDGMENTS   %%%%%%%%%%%%%%%%%%%%%%%%%%%%%%%%%%%%%%%%%%%%%%%%%%%
\begin{acknowledgments}
O. A. acknowledges support from the European Union (Project No. ITN-2010-264951, CORINF).
R. L. thanks Pr. F. Leyvraz for his hospitality at the "Centro Internacional de Ciencias", Cuernavaca, Mexico. 
\end{acknowledgments}

%%%%%%%%%%%%%%%%%%%%%%%%%%%%%%%%%%%%%%%%%%%%%%%%%%%%%%%%%%%%%%%%%%%%%%%%%%%%%%%%%%ù
\bibliographystyle{apsrev}

\end{document}